\documentclass[letter]{aa} 

\usepackage{natbib}

\usepackage{caption}
\usepackage{subcaption}
\usepackage{adjustbox}
\usepackage[flushleft]{threeparttable}

\usepackage{graphicx}
\usepackage{txfonts}
\usepackage{subfig}

\usepackage{gensymb}
\usepackage{amssymb}

\usepackage{url}

\makeatletter
\newcommand\footnoteref[1]{\protected@xdef\@thefnmark{\ref{#1}}\@footnotemark}
\makeatother

\usepackage{tikz}
\def\checkmark{\tikz\fill[scale=0.4](0,.35) -- (.25,0) -- (1,.7) -- (.25,.15) -- cycle;} 
\usepackage{amstext}

\usepackage{xcolor}

\begin{document}

   \title{Unveiling the origin of the radio emission in radio-quiet quasars}



   \author{N. Herrera Ruiz
          \inst{1}
          \and
          E. Middelberg\inst{1}
          \and
          R. P. Norris\inst{2,3}
          \and
          A. Maini \inst{2,4,5,6}
          }

   \institute{Astronomisches Institut, Ruhr-Universit\"at Bochum, 
              Universit\"atstrasse 150, 44801 Bochum, Germany\\
              \email{herrera@astro.rub.de}
             \and CSIRO Australia Telescope National Facility, PO Box 76, Epping, NSW 1710, Australia
             \and Western Sydney University, Locked Bag 1797, Penrith South, NSW 1797, Australia
             \and Dipartimento di Fisica e Astronomia, Universit\`a di Bologna, viale B. Pichat 6/2, 40127 Bologna, Italy
             \and Department of Physics and Astronomy, Macquarie University, Balaclava Road, North Ryde, NSW, 2109, Australia
             \and INAF-IRA, via P. Gobetti 101, 40129 Bologna, Italy
             }

   \date{Received February 12, 2016; accepted March 12, 2016} 

 
  \abstract
   {The origin of the radio emission in radio-quiet quasars (RQQs) has been a matter of debate for a long time. It is not well understood whether the emission is caused by star formation in the host galaxy or by black hole activity of the active galactic nuclei (AGN). We shed some light on these questions using the Very Long Baseline Interferometry (VLBI) technique to search for RQQs in the field of the Cosmological Evolution Survey (COSMOS). The extensive multi-wavelength coverage of the field (from radio to X-rays) was used to classify RQQs, and the milli-arcsecond resolution of VLBI provides a direct way to identify AGNs. In a sample of 18 RQQs we detected 3 using the Very Long Baseline Array (VLBA) at 1.4 GHz. In this letter we report for the first time on a sample of RQQs with a measured lower limit on the fraction of radio emission coming from the AGN, thus demonstrating that the radio emission of at least some RQQs is dominated by an AGN.}

   \keywords{galaxies: active --
                radio continuum: galaxies -- quasars: general -- galaxies: evolution              
               }

   \maketitle
%

\section{Introduction}
\label{sec:int}

The study of the physical processes that occur in quasars is fundamental to understanding galaxy evolution. Quasars are the most energetic active galactic nuclei (AGN). When their radio emission is considered, they are typically classified into radio-quiet
quasars (RQQ) and radio-loud quasars (RLQ). The radio emission in RLQ is associated with powerful relativistic jets, while the origin of the radio emission in RQQ has remained  a matter of debate for a long time. The two main scenarios ascribe the radio emission to either star-forming activity in the host galaxy \citep{padovani2011, bonzini2013} or non-thermal processes near supermassive black holes, which are  less powerful  than  RLQs, that is, a scaled-down version of the RLQ sources \citep{prandoni2010}. \citet{kimball2011} and \citet{condon2013} analysed the radio luminosity function of optically selected quasar samples and found that RQQs are mostly driven by star-forming activity in the host galaxy, whereas AGNs must dominate at the bright end. \citet{kimball2011} limited their sample to nearby quasars with low redshifts between 0.2 and 0.3 to reach radio luminosities lower than $10^{22}\,W\,Hz^{-1}$. \citet{condon2013} studied low- and high-redshift samples with 0.2\,<\,z\,<\,0.45 and 1.8\,<\,z\,<\,2.5, respectively. They obtained a median spectral luminosity L$_{1.4GHz}(W\,Hz^{-1})$\,$\approx$\,$10^{22.7}$ for their low-redshift quasar sample and L$_{1.4GHz}(W\,Hz^{-1})$\,$\approx$\,$10^{24.1}$ for their high-redshift quasar sample. The sources presented in this work are at moderate to high redshifts between 1.2 and 2.9 and their radio luminosities are in the range\,$\sim$\,$10^{24-25}\,W\,Hz^{-1}$. Regardless of the radio power, the strong redshift evolution of the luminosity function found by \citet{padovani2015} suggests that the dividing line between RQ and RL is redshift dependent and so would be higher at higher redshifts.

While RLQs dominate the population at radio flux densities higher than 1 mJy, they become rarer at lower radio flux densities. The flattening in the normalized source counts below 1 mJy has mostly been explained by the emergence of a faint population of star-forming galaxies. Nevertheless, multi-wavelength classifications provided evidence that RQQs still contribute significantly to the sub-mJy population \citep{smolcic2008, padovani2011}. 

\citet{dunlop2003} reported that the RQQs are mainly hosted by ellipticals. Identifying radio-emitting AGNs unambiguously is usually difficult since compact interferometers cannot distinguish radio emission coming from AGN or star formation. Very Long Baseline Interferometry (VLBI) observations provide milli-arcsecond-scale resolution, and a detection at 1.4\,GHz therefore requires brightness temperatures that can generally only be reached by the non-thermal emission processes in AGNs. A direct way to positively identify AGN in RQQs would be, in principle, to detect them using VLBI. In particular, the 5.5\,$\sigma$ sensitivity limit of the VLBA achieved by the observations from \citet{herreraruiz} is $55\,\mu$Jy, which corresponds to a brightness temperature of $3\times 10^{5}$\,K and a luminosity of $1.4\times 10^{21}$\,W/Hz (at a redshift of 0.1). \citet{kewley2000} found that compact radio cores with luminosities higher than $10^{21}$ W/Hz are almost certainly AGN. Occasionally this brightness temperature can also be reached by star-forming activity or radio supernovae. On the other hand, the transient events are extremely unusual at radio wavelengths. In particular, only $\sim$50 radio supernovae have been detected in the whole sky after $\text{about }$30 years of observations \citep{lien2011}.

Quasars show broad emission lines and are typically seen at high redshifts. They can be classified as RQQ through the ratios between their radio flux density and their optical, infrared, and X-ray flux densities (see Sect.~\ref{sec:cla}), for which comprehensive  multi-wavelength coverage is required (see Sect.~\ref{sec:anc}).

In this letter we report the detection of three RQQ with the Very Long Baseline Array (VLBA) (see Sect.~\ref{sec:dis}). The results presented here agree excellently well with the results obtained by \citet{maini}.

Throughout this letter, we adopt a flat $\Lambda$CDM cosmology with ${\it{H}}_{0}$ = 67.3 km s$^{-1}$ Mpc$^{-1}$, $\Omega_{M}$ = 0.31 and $\Omega_{\Lambda}$ = 0.69 (according to the recent Planck results published by \citealt{planck2014}).

\section{Sample}
\label{sec:sam}

\citet{herreraruiz} observed $\sim$3000 radio sources from \citet{schinnerer2010} in the COSMOS field with the VLBA at 1.4 GHz, obtaining milli-arcsecond resolution images at a 1$\sigma$ sensitivity of 10\,$\mu$Jy/beam in the central part of the field. The main goal of the project was to study the AGN component of the faint radio population in a statistical way (the field was chosen for the absence of strong sources). They have detected 468 radio sources with a signal-to-noise ratio higher than 5.5.

At a later stage of the project, we searched for RQQs on the \citet{schinnerer2010} COSMOS survey to investigate if any of them was detected with the VLBA. For that purpose, we cross-matched the radio catalogue from \citet{schinnerer2010} with the quasars catalogue from \citet{flesch2015}. Within a matching radius of 0.5\arcsec, 38 counterparts were identified. 

To ensure the quasar classification we considered the quasar definition derived from the absolute magnitude M$_{B}$\,<\,-23 given by \citet{schmidt1983}. For the 38 counterparts, we found that 24 sources satisfy this criterion.

To distinguish between RL and RQ quasars we used the infrared-to-radio ratio, $q_{24}$, and the radio-to-optical ratio, $R_{V}$, defined as (all logarithms are base-10)

\begin{itemize}
\renewcommand\labelitemi{--}
\item $q_{24}$=log($L_{24\mu m}/L_{1.4GHz}$), where $L_{24\mu m}$ and $L_{1.4GHz}$ are the luminosity of the source at 24 $\mu$m and 1.4 GHz, respectively.

\item $R_{V}$=log($S_{1.4GHz}$/$S_{V}$), where $S_{1.4GHz}$ and $S_{V}$ are the flux density of the source measured at 1.4 GHz and $V$-band, respectively.

\end{itemize}

We considered a quasar as RQ when $q_{24}$ > 0 \citep{appleton2004} and $R_{V}$ < 1.4 \citep{padovani2011} simultaneously, to minimize the probability of biasing the classification through the q24 criterion alone, since the 24\,$\mu$m luminosity could be contaminated by the AGN emission.

We cross-matched the 24 quasars with the Multiband Imaging Photometer for Spitzer (MIPS) data from NASA/IPAC Infrared Science Archive\footnote{\label{note1}\url{http://irsa.ipac.caltech.edu/Missions/spitzer.html}}
and with the photometric catalogue from \citet{capak2007}, using a matching radius of 0.5\arcsec in both cases. 

We found that 18 of the 24 sources fitted this RQ classification. \citet{herreraruiz} detected 3 of these 18 RQQs with the VLBA. This implies a detection rate of $0.17^{+0.12}_{-0.05}$. Uncertainties correspond to the 1\,$\sigma$ confidence intervals on binomial population proportions using the Bayesian beta distribution quantile technique as demonstrated by  \citet{cameron2011}.

We extensively studied the three VLBA sources at different wavelengths using additional criteria to ensure that they are classified
correctly as RQQs.


\section{Additional classification criteria for radio-quiet quasars}
\label{sec:cla}

\citet{hao2014} compared different radio-loudness criteria to classify the XMM-COSMOS quasar sample into radio loud (RL) and radio quiet (RQ) quasars. Here we use their criteria to double-check the classification of our three VLBA-detected RQQs (all logarithms are base-10):

\begin{itemize}
\renewcommand\labelitemi{--}
\item $R_{i}$=log($L_{1.4GHz}/L_{i}$), where $L_{1.4GHz}$ and $L_{i}$ are the luminosity of the source at 1.4 GHz and $i$ band. The quasar is classified as RQ when $R_{i}$ < 1 \citep{balokovic2012}.

\item $R_{x}$=log($\nu L_{\nu}(5GHz)/L_{x}$(2-10keV)), where $\nu L_{\nu}$(5GHz) and $L_{x}$(2-10keV) are the luminosity of the source at 5 GHz and X-ray hard (2-10 keV) band. The quasar is classified as RQ when $R_{x}$ < $-$3 \citep{terashima2003}.

\item $P_{5GHz}$=log($P_{5GHz}$(W/Hz/Sr)), where $P_{5GHz}$(W/Hz/sr) is the power of the source at 5 GHz. The quasar is classified as RQ when $P_{5GHz}$ < 24 \citep{goldschmidt1999}.

\end{itemize}

We also used the most widely used criterion in the literature (e.g. \citealt{banados2015}), the radio-to-optical flux density ratio defined by \citet{kellermann1989} as

\begin{itemize}
\renewcommand\labelitemi{--}

\item $R=S_{5GHz}/S_{4400\AA}$, where $S_{5GHz}$ is the 5GHz radio flux density and $S_{4400\AA}$ is the $4400\AA$ optical flux density. The quasar is classified as RQ when $R$ < 10.

\end{itemize}

\citet{padovani2011} used the following criterion to distinguish between RQ and RL AGNs and star-forming galaxies:

\begin{itemize}
\renewcommand\labelitemi{--}

\item The source is classified as RQQ when $R_{V}$ < 1.4 and $L_{x}$ > 10$^{35}$\,W, where $L_{x}$ is the luminosity of the source in the X-ray hard band (2-10 keV).

\end{itemize}

\citet{bonzini2013} classified a source as RQQ when it simultaneously
fulfilled the following two criteria:

\begin{itemize}
\renewcommand\labelitemi{--}

\item $q_{24obs}$ located within or above a locus defined by the M82 template. $q_{24obs}$ is defined as $q_{24obs}$=log($S_{24\mu m}/S_{1.4GHz}$), where $S_{24\mu m}$ and $S_{1.4GHz}$ are the 24\,$\mu$m and 1.4\,GHz observed flux densities. 

\item Source located within the ``Donley wedge'' \citep{donley2012} of the IRAC colour-colour diagram.

\end{itemize}


\section{Ancillary data}
\label{sec:anc}

The COSMOS field has been well studied, resulting in a great deal of ancillary data\footnote{\url{http://cosmos.astro.caltech.edu}}. We made use of the NASA/IPAC Extragalactic Database (NED) to recover information for the three RQQs at other wavelengths. We searched for counterparts, using as input ID the VLA name from the catalogue of \citet{schinnerer2010}.

\subsection{Spectroscopic redshift}

Thanks to the deep coverage, spectroscopic redshifts are available (from optical spectra) for all of the three sources detected with VLBA \citep{trump2009, civano2012}.

\subsection{Radio}

We converted the observed 1.4 GHz VLA radio flux densities from \citet{schinnerer2010} into 5 GHz assuming a radio spectral power law \mbox{$f_{\nu}$ $\propto$ $\nu^{\alpha}$}, where $f_{\nu}$ is the flux density measured at frequency $\nu$ and $\alpha$ is the spectral index. We adopted a spectral index $\alpha$=$-$0.75 (e.g. \citealt{kukula1998}).

We calculated the luminosity, in W/Hz, with the following expression: \mbox{$L = 4\pi D_{L} f_\nu$}, where $D_{L}$ is the luminosity distance and $f_\nu$ the measured flux density.

\subsection{Infrared}

We searched for infrared counterparts of our three sources in the Spitzer Enhanced Imaging Products (SEIP) source list, making use of the NASA/IPAC Infrared Science Archive\footnoteref{note1}. We found that each source has only one possible counterpart within a cross-matching radius of 4\arcsec, and that the separation is smaller than 0.5\arcsec in the three cases.

\subsection{Optical}

We cross-matched the three sources with the photometry catalogue
of \citet{capak2007} with a match radius of 0.5\arcsec. We found counterparts for each of them, with measured AB magnitudes in $B$-, $V$-, and $i$-band. To transform the AB magnitudes into $\mu$Jy flux densities, we made use of the expression {\mbox{$S$=$10^{c-AB/2.5}$}\,$\mu$Jy} \citep{oke1974}, where {\it {AB}} is the measured AB magnitude and c=9.56. 

We calculated the absolute magnitude using the following equation: M\,=\,m\,-\,5\,(log$_{10}$D$_{L}$\,-\,1), where m is the apparent magnitude and D$_{L}$ is the luminosity distance measured in parsecs.

\subsection{X-rays}

We cross-matched our three sources with the X-ray catalogue from \citet{brightman2014} choosing a matching radius of 1\arcsec, and found a counterpart for each of them. Both redshift and ID name used in the Chandra-COSMOS Survey \citep{elvis2009} coincide with the assigned spectroscopic redshift and the NED cross-ID of our sources.


\begin{table}
\centering
\caption{Classification of the three VLBA-detected RQQs with the different criteria.}             
\renewcommand{\arraystretch}{1.25}
\begin{adjustbox}{max width=0.48\textwidth}  
\label{table:check}     
\begin{threeparttable}
\centering                        
\begin{tabular}{l l l l l l | l l | l l}      
\hline\hline           
ID & q$_{24}$ & R$_{i}$ & R$_{x}$ & P$_{5GHz}$ & R & R$_{V}$ & L$_{x}$ & q$_{24obs}$ & DW$^{a}$ \\    
\hline                        
  C0366   &     \checkmark   &   \checkmark  &  $\times$   &   $\times$  &    \checkmark &  \checkmark   &   \checkmark  & \checkmark & \checkmark \\ 
  C1397   &    \checkmark   &   \checkmark  &  $\times$   &   \checkmark  &  \checkmark   &    \checkmark &   \checkmark  & \checkmark & \checkmark \\ 
  C1897    &  \checkmark   &  \checkmark  &    \checkmark &  \checkmark   &  \checkmark   &   \checkmark  &    \checkmark & \checkmark & \checkmark \\ 
\hline                                  
\end{tabular}
\begin{tablenotes}
\small
\item A checkmark (\checkmark) means the source is classified as RQQ, and a cross ($\times$) means is not. $^{a}$ Source located within the ``Donley Wedge''.
\end{tablenotes}
\end{threeparttable}
\end{adjustbox}
\end{table}

\subsection{False-ID rate}

\citet{herreraruiz} estimated the false-ID rate resulting from the cross-matching by shifting the positions of their VLA radio subsample by 1 arcmin (i.e. much larger than the VLA synthesized beam width). They then repeated the cross-matching procedure and found counterparts for 2 of their 468 shifted radio sources, giving a false-ID rate of around 0.4\%. Therefore, we are very confident that our VLBI detections correspond to the RQQ.


\section{Results}
\label{sec:dis}

In Table~\ref{table:check} we summarise the results of the classification criteria as defined in Sect.~\ref{sec:cla}. A checkmark (\checkmark) is used when a source fulfils a radio-quietness criterion and a cross ($\times$) when not. C1897 satisfies all criteria of radio-quietness, while C0366 and C1397 satisfy almost all of them. We are therefore confident that we are analysing the origin of the radio emission in RQQ.

Two of the VLBA-detected sources, C0366 and C1397, are unresolved, and C1897 is slightly resolved. The VLBA restoring beams for C0366, C1397, and C1897 are 15.9$\times$7.0 mas$^{2}$, 15.8$\times$7.3 mas$^{2}$ , and 17.2$\times$8.6 mas$^{2}$, respectively.

To evaluate if the radio emission of our VLBA-detected sources could be due to supernovae, we calculated the supernova rates needed for a young supernova ($\nu_{YSN}$) and for an old supernova remnant ($\nu_{SNR}$) to achieve the observed radio luminosity, following the equations from \citet{kewley2000}:
 
$L^{YSN}_{NT}\,=\,1.58\,\times\,10^{21}\,(\nu/8.44\,GHz)^{-0.74}\,\nu_{YSN}\,W\,Hz^{-1}$ 

$L^{SNR}_{NT}\,=\,1.77\,\times\,10^{22}\,(\nu/8.44\,GHz)^{-0.74}\,\nu_{SNR}\,W\,Hz^{-1}$, 

\noindent where $\nu$ is the frequency of observation. The obtained $\nu_{YSN}$ and $\nu_{SNR}$ values are 3342 and 298 $yr^{-1}$ (C0366), 1968 and 176 $yr^{-1}$ (C1397), and 402 and 36 $yr^{-1}$ (C1897), respectively. These values considerably exceed
the rates observed in starburst and ultra-luminous infrared galaxies (ULIRGs). In addition, star formation cannot reach the observed brightness temperatures, so the radio emission of our VLBA-detected sources must come from the AGN.

Table~\ref{table:rqq} lists the measured properties of the three RQQs. The column entries are:

\begin{itemize}
\renewcommand\labelitemi{--}
\item {\it {Col 1}}: Source ID, as used in \citet{herreraruiz}. 

\item {\it {Cols 2, 3}}: Right ascension and declination (in degrees, J2000), measured with the VLBA.

\item {\it {Col 4}}: spectroscopic redshift, from \citet{trump2009} for C0366 and C1897, and from \citet{civano2012} for C1397.

\item {\it {Col 5}}: Integrated 1.4 GHz VLA radio flux density (in $\mu$Jy), from \citet{schinnerer2010}.

\item {\it {Col 6}}: Integrated 1.4 GHz VLBA radio flux density (in $\mu$Jy).

\item {\it {Col 7}}: Signal-to-noise ratio obtained with the VLBA.

\item {\it {Col 8}}: Spitzer/MIPS 24 $\mu$m flux density (in $\mu$Jy), from NASA/IPAC Infrared Science Archive\footnoteref{note1}.

\item {\it {Col 9}}: K-corrected radio luminosity (in W/Hz), computed using the integrated 1.4 GHz VLA radio flux density.

\item {\it {Col 10}}: Lower limit of the radio brightness temperature (in K), computed using the integrated 1.4 GHz VLBA radio flux density.

\item {\it {Col 11}}: Base-10 logarithm of the rest-frame X-ray hard band (2-10 keV) luminosity (in W), from \citet{brightman2014}.

\item {\it {Cols 12, 13 and 14}}: $B$-, $V$-, and $i$-band magnitudes (AB system), from \citet{capak2007}.

\item {\it {Col 15}}: $B$-band absolute magnitude (AB system).

\item {\it {Col 16, 17, 18 and 19}}: Spitzer/IRAC 3.6, 4.5, 5.8 and 8.0 $\mu$m flux densities from NASA/IPAC Infrared Science Archive\footnoteref{note1}.

\end{itemize}

\begin{table*}
\caption{Properties of the three VLBA-detected RQQs.}
\label{table:rqq}
\begin{threeparttable}
\centering
\renewcommand{\arraystretch}{1.25}
\begin{adjustbox}{max width=\textwidth}
\begin{tabular}{lllllllllllllllllll} 
\hline\hline             
ID    &      RA    &   Dec    & z$_{sp}$ & $S_{VLA}$ & $S_{VLBA}$ & S/N  & $S_{24\mu m}$ & log$L_{1.4}$ & T$_{B}$ & logL$_{x}$ &    $V$   &   $B$    &   $i$  & M$_{B}$ & $S_{3.6\mu m}$ & $S_{4.5\mu m}$ & $S_{5.8\mu m}$ & $S_{8.0\mu m}$ \\
      &    (deg)   &   (deg)  &          & ($\mu$Jy) &  ($\mu$Jy) &      &   ($\mu$Jy)   &  (W/Hz)  &  ($\times 10^{5}$K)  &   (W)      & (mag) & (mag) & (mag) & (mag) & ($\mu$Jy) & ($\mu$Jy) & ($\mu$Jy) & ($\mu$Jy) \\
(1)   &     (2)    &   (3)    &   (4)    &    (5)    &     (6)    & (7)  &      (8)      &         (9)       &  (10)   &  (11)    &   (12)  & (13) & (14)  & (15) & (16) & (17) & (18)  & (19) \\
\hline
C0366 & 149.649932 & 1.865855 &   2.94   &   362     &     255    & 16.4 &      658      &  25.3  &  >11.8   &   37.01       &   20.04  &   20.02  &  19.73 &  -26.9  &  120 & 128 & 169 & 276 \\   
C1397 & 150.037183 & 2.244620 &   2.54   &   293     &     156    & 8.2  &     2904      &  25.07  &  >7.0   &   36.47       &   19.9   &   20.4   &  19.2  &  -26.2  & 156  & 229 & 406 & 776 \\
C1897 & 150.207967 & 2.083336 &   1.24   &   311     &     236    & 7.5  &     3052      &   24.38  &  >8.2   &  37.7        &   20.1   &   20.4   &  19.4 & -24.3  & 259  & 422 & 657 & 1039 \\ 
\hline
\end{tabular}
\end{adjustbox}
\end{threeparttable}
\end{table*}

\subsection{Individual sources}

\subsubsection{C0366}

C0366 has been classified as a broad absorption-line quasar by \citet{trump2006} and \citet{scaringi2009} and has a spectroscopic redshift of 2.94 \citep{trump2009}. It is a faint radio source with an integrated VLA radio flux density of 362 $\mu$Jy at 1.4 GHz. Our VLBA observation recovered $\sim$70\% of this VLA radio flux density. This means that a large amount of the radio emission comes from the AGN activity. Moreover, because of the high resolution achieved with the VLBA, this has to be treated as a lower limit. The signal-to-noise ratio (S/N) obtained with the VLBA is 16.4.

\subsubsection{C1397}

C1397 has been classified as a quasar by \citet{richards2009} and \citet{stern2012} and has a spectroscopic redshift of 2.54 \citep{civano2012}. It is a faint radio source with an integrated VLA radio flux density of 293 $\mu$Jy at 1.4 GHz. Only about half of the radio flux density measured with the VLA is detected with the VLBA, suggesting that at least half of the radio emission comes from the AGN activity. Nevertheless, because of the high resolution achieved by the VLBA, this has to be considered a lower limit. The S/N obtained with the VLBA is 8.2.

\subsubsection{C1897}

C1897 has been classified as a quasar by \citet{richards2009} and \citet{veron2010} and has a spectroscopic redshift of 1.24 \citep{trump2009}. It is a faint radio source with an integrated VLA radio flux density of 311 $\mu$Jy at 1.4 GHz. The radio flux density measured with the VLBA at 1.4 GHz is 75\% of that measured with the VLA. This means that the radio emission comes predominantly from the AGN activity. Again, this has to be considered a lower limit. The S/N obtained with the VLBA is 7.5.

\section{Conclusions}
\label{sec:con}

From a sample of 18 RQQs in the COSMOS field, we detected 3 using the VLBA with a high S/N. This implies a detection rate of $0.17^{+0.12}_{-0.05}$. For the remaining 15 we cannot make a statement. The extensive ancillary data on the COSMOS field allowed us to reinforce their classification. We used seven different criteria to discern between radio-quiet and radio-loud quasars. C1897 is RQ according to all the seven criteria, while C0366 and C1397 are RQ according to most of them. The three RQQs lie at moderately high redshifts (between 1.2 and 3) and their integrated VLA radio flux densities at 1.4 GHz are of the order of 300\,$\mu$Jy. 

The number of RQQs detected (3 of 18) together with their radio flux densities (well above 100\,$\mu$Jy) agree well with the suggestion that the radio emission of RQQ cores provides a relevant contribution to the sub-millijansky radio flux population \citep{smolcic2008, padovani2011}.

The VLBA-measured radio flux densities are between 50\% and 75\% of the VLA radio flux densities. This shows that at least in some RQQs the radio emission mainly comes from non-thermal AGN activity and not from star-forming processes, making the former scenario more relevant. \citet{maini} have obtained a similar result on a different sample, reinforcing the conclusions presented here.

\vspace{0.5cm}

\begin{acknowledgements}
      NHR acknowledges support from the Deutsche Forschungsgemeinschaft through project MI 1230/4-1. We wish to thank the referee Ken Kellermann for the
valuable comments, which have helped to improve this letter considerably. This research made use of \texttt{Topcat} \citep{taylor2005}, available at \url{http://www.starlink.ac.uk/topcat/}. This research has made use of the NASA/IPAC Extragalactic Database (NED), which is operated by the Jet Propulsion Laboratory, California Institute of Technology, under contract with the National Aeronautics and Space Administration. The VLBA is operated by the National Radio Astronomy Observatory, a facility of the National Science Foundation operated under cooperative agreement by Associated Universities, Inc.
\end{acknowledgements}


\setlength{\bibsep}{0pt plus 0.3ex}
\bibliographystyle{agsm}
\bibliography{references}

\end{document}